\documentstyle[12pt]{article}
\begin{document}
\large
\begin{center}
{\bf Trilinear representation and the Moutard transformation for
the Tzitz\'{e}ica equation}\\
\vspace{7mm}

O.V.Kaptsov, Yu.V.Shan'ko
\end{center}

\begin{center}
{\small Computing Center, Academy of Sciences,\\
  Academgorodok, 660036,Krasnoyarsk, Russia\\
E-mail: kaptsov@cckr.krasnoyarsk.su}
\end{center}

\begin{abstract}
In this paper we present a trilinear form and a Darboux-type
transformation to equation
$$
(\ln v)_{xy} = v - 1/v^2
$$
considered by Tzitz\'{e}ica in 1910. Soliton solutions are constructed
by dressing the trivial solution.
\end{abstract}

\noindent
{\bf 1. Introduction.}
 In this work we construct exact solutions to the equation
$$  (\ln v)_{xy} = v - 1/v^2 . \eqno(1.1)$$
This equation originated in differential geometry \cite{tz1,tz2} as
the compatibility condition of the linear system
$$\theta_{xy} = v\theta, \quad \theta_{xx} = (v_x \theta_x + \theta_y)/v,
\quad  \theta_{yy} = (v_y \theta_y + \theta_x)/v.   \eqno(1.2) $$
This allows us to call $v$ the potential and (1.1) the
Tzitz\'{e}ica equation. Any three linearly independent solutions of (1.2)
generate a surface in $R^3(x^1,x^2,x^3)$
      $$ x^i = \theta^i(x,y), \quad i=1,2,3,   $$
which is called T-surface.  Tzitz\'{e}ica discovered a relation
between T-surfaces. He established that the map
$$\theta^i \ \longrightarrow \  \theta^i - (1+c)(\ln r)_{y} \theta^i_x /v -
 (1-c)(\ln r)_{x} \theta^i_y /v, \eqno(1.3)$$
$$ i = 1, 2, 3,$$
where $c$ is an arbitrary constant and $r$ is a solution of the system
$$  r_{xy} = vr, \quad r_{xx} = (v_x r_x + \lambda r_y)/v, \quad
r_{yy} = (v_y r_y + \lambda^{-1} r_x)/v,   \eqno(1.4)  $$
$$\lambda = (c+1)/(c-1) \ , $$
transforms T-surface into a new T-surface which corresponds to the potential
$$  v^{\prime} = - v + 2(\ln r)_x (\ln r)_y. \eqno (1.5) $$
It is easy to show that equation (1.1) is the compatibility condition of
the linear system  (1.4)

Setting $v = exp(u)$, the Tzitz\'{e}ica equation becomes
Bullough-Dodd-Jiber-Shabat one
$$
u_{xy} = \exp(u) - \exp(-2u) . \eqno (1.6)
$$
The "zero curvative" representation  for (1.6) has a form  \cite{mi}
$$\psi_x = L \psi , \qquad \psi_y = A \psi , \eqno (1.7) $$
where $\psi = (\psi_1,\psi_2,\psi_3)$ is a vector function, $L$ and $A$ are
the matrices
$$L=\left(\matrix{-u_x & 0 & \lambda \cr
                   1   & u_x & 0  \cr
                   0   & 1   & 0  \cr  }\right) , \qquad
A=\left(\matrix{   0   & e^{-2u} & 0 \cr
                   0   & 0       & e^{u}  \cr
       \lambda^{-1}e^u   & 0   & 0  \cr  }\right)   .
$$
If we replace the vector function $\psi$ by $(\lambda r_y/v, r_x, r)$, then
it is possible to derive from (1.7) the representation (1.4)  and vice versa.
The finite-gap solutions  and B\"acklund transformation to equation (1.6)
can be found in \cite{sh1,sh2}. As shown in \cite{kap}, some solutions of
eq.(1.6) can be expressed in terms of elliptic functions .
However, an explicit formula for N-soliton solutions does not, to our
knowledge, appear in the literature.

One of the helpful tools in the study of integrable equations, over the past
25 years, has been the Hirota bilinear formalism \cite{hir} (multilinear
operators can be found in \cite{gram}). In recent publication  we use
a trilinear equation which allows us to derive multiparameter solutions
of (1.1). We introduce a special case of the Moutard transformation
\cite{dar,nim} to the system (1.4) and provide the proof of the N-soliton
formula.

{\bf 2. $\tau$-functions and explicit solutions.}

In order to obtain the trilinear form of equation (1.1), we carry out the
following change of variables:
$$v=1 - 2(\ln\tau)_{xy} . \eqno(2.1)$$
Note that, an analogous change of variables for equation
$$ \bigtriangleup (\ln w) = w - 1/w^2$$
was used in \cite{kap2}. Substitution of (2.1) into (1.1) yields
$$(1-2(\ln\tau)_{xy})^2 [(\ln(\tau^2-2\tau\tau_{xy}+2\tau_{x}\tau_{y}))_{xy}
-1] + 1 = 0 \eqno(2.2).$$
Multiplying (2.2) by $\tau^3$, we obtain the trilinear equation.
It turns out that, the equation (2.2) has N-soliton solutions of the form
$$1 +  \sum \limits_{i = 1}^N f_i +
\sum \limits_{k = 2}^N \sum \limits_{1\leq i_1 <\cdots < i_k \leq N}
c_{i_1 \cdots i_k}f_{i_1}\cdots f_{i_k}, $$
where
$$f_i = \exp(k_i x + 3 y/k_i +s_i), \eqno(2.3)$$
 $k_i, s_i$ are arbitrary constants and  $c_{i_1 \cdots i_k}$ are constants
to be determined. One can show by straightforward analysis that
the one-soliton solution of (2.2) takes the form
$$\tau_1 = 1 + exp(kx + 3y/k + s),$$
with $k$ and $s$ being arbitrary constants.

To find two-soliton solution, we insert the function
$$\tau_2 = 1 + f_1 + f_2 + p_{12} f_1 f_2 , \eqno (2.4) $$
into (2.2) and obtain a polynomial equation with respect to $f_1, f_2$. Since
the coefficients of $f_1, f_2$ must vanish, we get many equations
with respect to $p_{12}.$ Nevertheless,
$$p_{12} = p(k_1,k_2) \equiv \frac{(k_1-k_2)^2 (k_1^2-k_1k_2+k_2^2)}
{(k_1+k_2)^2 (k_1^2+k_1k_2+k_2^2)}  \eqno(2.5)$$
satisfies all these equations.

Now we can suppose that N-soliton solution is
$$
\tau_N = 1 + \sum_{i=1}^{N} f_i + \sum_{m=2}^{N}
(\sum \limits_{1\leq i_1 <\dots <i_m \leq N} f_{i_1}\cdots f_{i_m}
\prod \limits_{1\leq j < r \leq m} p(k_{i_j},k_{i_r})), \eqno(2.6)
$$
where $p(k_{i_j},k_{i_r})$ are given by (2.5). It can be checked by
straightforward calculations that $\tau_3$ satisfies (2.2). In the section 4
we shall prove that  $\tau_N$ is a solution of (2.2).

It is rather interesting to find solutions which are expressed in the terms
of elementary functions but differ from N-solitons. The corresponding solutions
can be found using linear differential constraints \cite{kap2}. In our case,
these constraints are given by ordinary differential equations
$$\partial _x\prod \limits_{1\leq m\leq N}(\prod \limits_{1\leq
i_1<...<i_m\leq N} (\partial _x-k_{i_1}-k_{i_2}-...-k_{i_m}))\tau = 0,
$$
$$\partial _y\prod \limits_{1\leq m\leq N}(\prod \limits_{1\leq
i_1<...<i_m\leq N} (\partial _y-3/k_{i_1}-3/k_{i_2}-...-3/k_{i_m}))\tau = 0,
$$
where $\partial _x = \frac{\partial}{\partial x}$,
$\partial _y = \frac{\partial}{\partial y}.$ If $N=2$, then the previous
equations take the form
$$\partial _x(\partial _x - k_1)(\partial _x - k_2)(\partial _x - k_1 - k_2)
\tau = 0 , \eqno(2.7)$$
$$\partial _y(\partial _y - 3/k_1)(\partial _y - 3/k_2)
(\partial _y - 3/k_1 - 3/k_2)\tau = 0 . \eqno(2.8)$$
In particular, we consider two short examples.

\noindent
{\it Example 1.}
Suppose that $k_1 = k_2 = k \neq 0$, then the function
$$1 + c(x-3y/k^2)\exp(kx + 3y/k) - \frac{c^2}{12k^2}\exp(2kx+6y/k),
\qquad c\in R$$
satisfies (2.2).

\noindent
{\it Example 2.}
Assuming that $k_1 = ib$, $k_2  = -ib$, one can find the solution
$$ \sin(bx-3y/b) + \sqrt{3}(bx+3y/b).$$

{\bf 3. Moutard-Tzitz\'{e}ica transformation.}
In this section we describe another way of constructing solutions to
equation (1.1). This approach is based on the so-called Darboux transformation
\cite{mat}. It should be noted that Moutard discovered the corresponding
transformation in 1870 \cite{dar,nim}.

One may define the classical Moutard transformation in the following manner.
Let $r$ and $r_1$ be linearly independent solutions of the equation
 $$R_{xy} = v(x,y)R , \eqno(3.1) $$
then the function $r^{\prime}$ satisfying the system
$$(r_1 r^{\prime})_x = r_1^2(r/r_1)_x ,
\qquad (r_1 r^{\prime})_y = -r_1^2(r/r_1)_y, \eqno(3.2) $$
is a solution of the equation
 $$R^{\prime}_{xy} = v^{\prime}(x,y)R^{\prime} , \eqno(3.3)$$
where
$$v^{\prime} = v - 2(\ln r_1)_{xy}. \eqno(3.4)$$

It turns out that there exists a special case of the Moutard transformation
for the system (1.4). We omit all calculations and present the final formula
$$ r^{\prime} =
 \frac{-2\lambda_1 r_{1_y} r_x + 2\lambda r_{1_x} r_y +
(\lambda_1-\lambda)v r r_1}{- (\lambda_1+\lambda ) v r_1} . \eqno (3.5)$$
Here $r$, $r_1$ are solutions of the system (1.4) (with potential $v$)
corresponding to different $\lambda$, $\lambda_1$ and $r^{\prime}$ satisfies
(1.4) with the same $\lambda$ but the modified potential (3.4).
Describing this transformation, we shall say that $r^{\prime}$ is obtained
from $r$ by $r_1$ and call (3.5) the Moutard-Tzitz\'{e}ica transformation
(or MT-transformation).

It can be shown that the Moutard-Tzitz\'{e}ica transformation has the
property of permutability . To define more exactly this property it is
necessary to introduce some notations.
Denote by $r^0(\lambda)$ a solution of the system (1.4) with the potential
$v^0$. Let $S$ be the skew-symmetric map
$$S(\mu,\lambda) = \frac{2\mu r^0(\mu)_y r^0(\lambda)_x -
2\lambda r^0(\mu)_x r^0(\lambda)_y + (\lambda - \mu)v^0r^0(\mu)r^0(\lambda)}
{(\mu + \lambda) v^0}\eqno(3.6)$$
Then the solution $r^1(\lambda_1,\lambda)$ obtained from $r^0(\lambda)$
by $r^0(\lambda_1)$ takes the form
$$r^1(\lambda_1,\lambda) =
S(\lambda_1,\lambda)/r^0(\lambda_1) . \eqno(3.7)$$
This solution corresponds to the potential
$$v^1(\lambda_1) =  v^0 - 2(\ln r^0(\lambda_1))_{xy} .\eqno(3.8) $$
We can apply the MT-transformation once again and obtain a new solution
$r^2(\lambda_1,\lambda_2,\lambda)$ from $r^1(\lambda_1,\lambda)$ by means
of $r^1(\lambda_1,\lambda_2).$ The solution $r^2(\lambda_1,\lambda_2,\lambda)$
corresponds to the potential
$$  v^2(\lambda_1,\lambda_2) = v^1(\lambda_1) -
        2(\ln r^1(\lambda_1,\lambda_2))_{xy} .  $$
On the other hand, we can first obtain the solution $r^1(\lambda_2,\lambda)$
from $r^0(\lambda)$  by  $r^0(\lambda_2)$ and then transform this solutions by
means of $r^1(\lambda_2,\lambda_1).$ It leads to the solution
$r^2(\lambda_2,\lambda_1,\lambda)$ corresponding to the potential
$$  v^2(\lambda_2,\lambda_1) = v^1(\lambda_2) -
        2(\ln r^1(\lambda_2,\lambda_1))_{xy} .  $$
The property of permutability means that
    $$v^2(\lambda_1,\lambda_2) = v^2(\lambda_2,\lambda_1),$$
$$r^2(\lambda_1,\lambda_2,\lambda) = r^2(\lambda_2,\lambda_1,\lambda).$$
{\bf Lemma 1.} The functions
$v^2(\lambda_1,\lambda_2)$, $r^2(\lambda_1,\lambda_2,\lambda)$ are invariant
under the transformation
$\lambda_1 \rightarrow \lambda_2$, $\lambda_2 \rightarrow \lambda_1$
and can be written as
$$v^2(\lambda_1,\lambda_2) = v^0 - (\ln S(\lambda_1,\lambda_2)^2)_{xy} ,
                    \eqno(3.9) $$
$$r^2(\lambda_1,\lambda_2,\lambda) =
\frac{r^0(\lambda) S(\lambda_1,\lambda_2)+
  r^0(\lambda_1) S(\lambda_2,\lambda)-
  r^0(\lambda_2) S(\lambda_1,\lambda)}{S(\lambda_1,\lambda_2)}.\eqno(3.10)
$$
{\it Proof:} It is easy to see that the relations (3.6),(3.7) and (3.8)
lead to
$$v^2(\lambda_1,\lambda_2) = v^0 - 2(\ln r^0(\lambda_1))_{xy} -
2(\ln S(\lambda_1,\lambda_2)/r^0(\lambda_1))_{xy} = $$
$$= v^0 - (\ln S(\lambda_1,\lambda_2)^2)_{xy} .$$
The equality (3.10) follows from the theorem of permutability for the
Moutard transformation \cite{bian}.

The property of permutability can be represented by means of the commuting
diagram

\begin{picture}(270,120)
\put(80,75){\vector(2,1){40}}
\put(200,95){\vector(2,-1){40}}
\put(80,55){\vector(2,-1){40}}
\put(200,35){\vector(2,1){40}}
\put(30,60){\makebox{ $(v^0,r^0(\lambda))$}}
\put(220,60)
{\makebox{ $(v^2(\lambda_1,\lambda_2),r^2(\lambda_1,\lambda_2,\lambda))$}}
\put(100,100){\makebox{ $(v^1(\lambda_1),r^1(\lambda_1,\lambda))$}}
\put(100,10){\makebox{ $(v^1(\lambda_2),r^1(\lambda_2,\lambda))$}}
\end{picture}

It is possible to apply the MT-transformation for constructing solutions
$r^3(\lambda_1,\lambda_2,\lambda_3,\lambda)$,\dots,
$r^n(\lambda_1,\dots,\lambda_n,\lambda)$ with the corresponding potentials
$v^3(\lambda_1,\lambda_2,\lambda_3)$,\dots,$v^n(\lambda_1,\dots,\lambda_n)$.
The recurrence relations for these functions are
  $$r^{n+1}(\bar{\lambda},\lambda_{n+1},\lambda) = \frac
{S^{n}(\bar{\lambda},\lambda_{n+1},\lambda)}{r^n(\bar{\lambda},\lambda_{n+1})}
\eqno(3.11)$$
$$v^{n+1}(\bar{\lambda},\lambda_{n+1}) = v^n(\bar{\lambda}) -
  2(\ln r^n(\bar{\lambda},\lambda_{n+1}), \eqno(3.12) $$
where $\bar{\lambda} = (\lambda_1,\dots,\lambda_n)$, and
$S^{n}(\bar{\lambda},\lambda_{n+1},\lambda)$ is defined by
$$  \frac{
2\lambda_{n+1}r^n(\bar{\lambda},\lambda_{n+1})_y r^n(\bar{\lambda},\lambda)_x
 - 2\lambda r^n(\bar{\lambda},\lambda_{n+1})_x r^n(\bar{\lambda},\lambda)_y}
{(\lambda + \lambda_{n+1})v^n(\bar{\lambda})}  +$$
$$+ (\lambda - \lambda_{n+1})r^n(\bar{\lambda},\lambda_{n+1})
r^n(\bar{\lambda},\lambda)/(\lambda + \lambda_{n+1}) . \eqno(3.13) $$
The following theorem holds:\\
{\bf Theorem 1.}
The functions  $(v^n(\lambda_1,\dots,\lambda_n)$,
 $r^n(\lambda_1,\dots,\lambda_n,\lambda)$ are invariant under an arbitrary
permutation of $\lambda_1,\dots,\lambda_n$.\\
{\it Proof by induction.}  For $n=2$ it follows from Lemma 1. If we assume
the Theorem to be true up to $n$, then owing to (3.11)-(3.13), the functions
$v^{n+1}(\lambda_1,\dots,\lambda_{n+1})$,
$r^{n+1}(\lambda_1,\dots,\lambda_{n+1},\lambda)$ are invariant under an
arbitrary permutation of $\lambda_1,\dots,\lambda_n$. In order to prove
the Theorem it is sufficient to establish that
 $$v^{n+1}(\tilde{\lambda},\lambda_n,\lambda_{n+1}) =
  v^{n+1}(\tilde{\lambda},\lambda_{n+1},\lambda_n), $$
$$r^{n+1}(\tilde{\lambda},\lambda_n,\lambda_{n+1},\lambda) =
  r^{n+1}(\tilde{\lambda},\lambda_{n+1},\lambda_n,\lambda),$$
where $\tilde{\lambda} = (\lambda_1,\dots,\lambda_{n-1}).$ The last formulae
follow from the obvious analogues of the (3.9), (3.10):
$$v^{n+1}(\tilde{\lambda},\lambda_n,\lambda_{n+1}) = v^{n-1}(\tilde{\lambda})
(\ln S^{n}(\bar{\lambda},\lambda_n,\lambda_{n+1}) )_{xy},  $$

$$ r^{n+1}(\tilde{\lambda},\lambda_n,\lambda_{n+1},\lambda) =
r^{n-1}(\tilde{\lambda},\lambda) \quad +$$
$$
+ \frac
{r^{n-1}
(\tilde{\lambda},\lambda_n) S^{n}(\tilde{\lambda},\lambda_{n+1},\lambda) -
r^{n-1}
(\tilde{\lambda},\lambda_{n+1}) S^{n}(\tilde{\lambda},\lambda_n,\lambda) }
{S^{n}(\tilde{\lambda},\lambda_n,\lambda_{n+1})}.
$$

As noted above, if $\tau^0$ is a solution of the equation (2.2), then the
function
 $$v^0 = 1 - 2 (\ln \tau^0)_{xy}$$
satisfies to the Tzitz\'{e}ica equation (1.1).
The iterated $\tau$-functions are defined as:
$$\tau^n(\lambda_1,\dots,\lambda_n) =
r^{n-1}(\lambda_1,\dots,\lambda_n)\tau^{n-1}(\lambda_1,\dots,\lambda_{n-1}),
\eqno(3.14)$$
where $r^{n-1}(\lambda_1,\dots,\lambda_n)$  given by (3.11).

\noindent
{\bf Lemma 2.}  The iterated $\tau$-functions satisfy the following relations
$$v^{n}(\lambda_1,\dots,\lambda_n) = 1 -
  2(\ln \tau^n(\lambda_1,\dots,\lambda_n) )_{xy}.\eqno(3.15) $$
{\it Proof.}  For $n=1$ we have
$$v^1(\lambda_1) = v^0 - 2(\ln r^0(\lambda_1))_{xy} = 1 - 2(\ln\tau^0)_{xy} -
2(\ln r^0(\lambda_1))_{xy}. $$
If we assume (3.15) to be true up to $n$, then from (3.12), one can obtain
$$v^{n+1}(\lambda_1,\dots,\lambda_{n+1}) = v^{n}(\lambda_1,\dots,\lambda_n) -
2(\ln r^n(\lambda_1,\dots,\lambda_n,\lambda_{n+1}) )_{xy} = $$
$$= 1 - 2(\ln \tau^n(\lambda_1,\dots,\lambda_n) )_{xy} -
2(\ln r^n(\lambda_1,\dots,\lambda_n,\lambda_{n+1}) )_{xy} = $$
$$= 1 -
2(\ln \tau^n(\lambda_1,\dots,\lambda_n)r^n(\lambda_1,\dots,\lambda_n,
\lambda_{n+1} )_{xy}=$$
$$
= 1 - 2(\ln \tau^{n+1}(\lambda_1,\dots,\lambda_{n+1}) )_{xy}.
$$
{\bf Lemma 3.} The function $\tau^n(\lambda_1,\dots,\lambda_n)$ is a skew-
symmetric map, i.e.,
$$\tau^n(\lambda_1,\dots,\lambda_i,\dots,\lambda_j,\dots,\lambda_n) =
- \tau^n(\lambda_1,\dots,\lambda_j,\dots,\lambda_i,\dots,\lambda_n),
\quad \forall i,j .$$
{\it Proof.} As a consequence of (3.7) and (3.14), one easily computes
$$\tau^1(\lambda_1) = r^0(\lambda_1)\tau^0,$$
$$\tau^2(\lambda_1,\lambda_2) = r^1(\lambda_1,\lambda_2)\tau^1(\lambda_1) =
 \frac{S(\lambda_1,\lambda_2)}{r^0(\lambda_1)}\tau^1(\lambda_1) =
 S(\lambda_1,\lambda_2)\tau^0. \eqno(3.16) $$
Since  $S(\lambda_1,\lambda_2)$ is a skew-symmetric function, $\tau^2$ is
a skew-symmetric map.
Assuming $\tau^n(\lambda_1,\dots,\lambda_n)$ is skew-symmetric, we will prove
that
$\tau^{n+1}(\lambda_1,\dots,\lambda_n,\lambda_{n+1})$ is
skew-symmetric too. According to (3.14), the function
$\tau^{n+1}(\lambda_1,\dots,\lambda_n,\lambda_{n+1})$ is skew-symmetric with
respect to $\lambda_1,\dots,\lambda_n$. Denote by $\tilde{\lambda}$ the
vector $(\lambda_1,\dots,\lambda_{n-1})$. It suffices to prove that
$$\tau^{n+1}(\tilde{\lambda},\lambda_n,\lambda_{n+1}) =
  - \tau^{n+1}(\tilde{\lambda},\lambda_{n+1},\lambda_n). $$
From (3.13) it follows that  $S^{n}(\tilde{\lambda},\lambda_n,\lambda_{n+1})$
is skew-symmetric with respect to $\lambda_n, \lambda_{n+1}$.
Since there exist obvious analogs of (3.16) for  $\tau^{n+1}$, we obtain
$$\tau^{n+1}(\tilde{\lambda},\lambda_n,\lambda_{n+1}) =
S^{n}(\tilde{\lambda},\lambda_n,\lambda_{n+1})\tau^{n-1}(\tilde{\lambda}) =$$
$$=- S^{n}(\tilde{\lambda},\lambda_{n+1},\lambda_n)\tau^{n-1}(\tilde{\lambda}
) = - \tau^{n+1}(\tilde{\lambda},\lambda_{n+1},\lambda_n).$$

We will write the map $\tau^n(\lambda_1,\dots,\lambda_n)$ as:
$\tau^1(\lambda_1)\wedge\cdots\wedge\tau^1(\lambda_n).$ It turns out that
this map is multilinear, i.e.,
$$\tau^1(\lambda_1)\wedge\cdots\wedge
( a\tau^1_1(\lambda_i)+ b\tau^1_2(\lambda_i) )\wedge\cdots \tau^1(\lambda_n)
=$$
$$= a\tau^1(\lambda_1)\wedge\cdots\wedge
\tau^1_1(\lambda_i)\wedge\cdots \tau^1(\lambda_n) +
b\tau^1(\lambda_1)\wedge\cdots\wedge
\tau^1_2(\lambda_i)\wedge\cdots \tau^1(\lambda_n),$$
     $$\forall a,b \in C , \quad  \forall i=1,\dots,n .$$
The basic properties of the map
$\tau^1(\lambda_1)\wedge\cdots\wedge\tau^1(\lambda_n)$ are stated below.

\noindent
{\bf Theorem 2.} The map
$\tau^n = \tau^1(\lambda_1)\wedge\cdots\wedge\tau^1(\lambda_n)$
is multilinear and skew-symmetric.\\
{\it Proof.} By Lemma 3 it suffices to prove that
$\tau^1(\lambda_1)\wedge\cdots\wedge\tau^1(\lambda_n)$  is a multilinear map.
Since the map is skew-symmetric, it is enough to establish the linearity of
$\tau^n$ with respect to $\tau^1(\lambda_n).$ This statement is proved by
induction.
Obviously, $r^1(\lambda_1,\lambda_2)$ is a linear function of $r^0(\lambda_2)$.
Suppose that
$r^n(\lambda_1,\dots,\lambda_{n+1})$ is a linear function with
respect to $r^0(\lambda_{n+1})$. Then it follows from (3.11) and (3.13)
that $r^{n+1}(\lambda_1,\dots,\lambda_{n+2})$ is a linear function of
 $r^0(\lambda_{n+2})$.

\noindent
{\bf Remarks:} We derived the Moutard-Tzitz\'{e}ica transformation
assuming $\lambda_1 \neq \lambda.$ In the case of $\lambda_1 = \lambda$,
we can introduce the following transformation
$$r^{\prime} = r - \frac{-2 r_x (\lambda r_{y\lambda} + r_y) +
  2 \lambda r_y r_{x \lambda} }{v r} . \eqno (3.17)$$
It can be checked that (3.17) transforms solutions of (1.4) (with the
potential $v$) in solutions of (1.4) with the modified potential
$$v^{\prime} = v - 2(\ln r)_{xy}.$$

\noindent
{\bf Verification of $N$-soliton formula.}
In this section we will obtain $N$-soliton solution by means of the
Moutard-Tzitz\'{e}ica transformation. For this purpose, we introduce a new
parameter $a$ by $\lambda = a^3$ and rewrite the Moutard-Tzitz\'{e}ica
transformation as
$$r^1(a_1,a) =
\frac{\Theta^0(r^0(a_1),r^0(a))}{r^0(a_1)},$$
where $\Theta^0(r^0(a_1),r^0(a))$ is given by
 $$\frac{2 r^0(a_1)_{xx} r^0(a)_x - 2 r^0(a_1)_x r^0(a)_{xx} +
(a^3 - a^3_1)r^0(a_1)r^0(a)}
{(a^3 + a^3_1) }. \eqno(4.1)$$
It follows from the system (1.4) that the function (4.1) is equivalent to
(3.6).

The iterated functions $r^n(a_1,\dots,a_n,a)$ are defined by recurrence
relations
$$r^n(a_1,\dots,a_n,a) =
\frac{\Theta^{n-1}(r^{n-1}(\a_1,\dots,a_n),r^{n-1}(a_1,\dots,a_{n-1},a))}
{r^{n-1}(a_1,\dots,a_n)}. \eqno(4.2)$$
Here the function $\Theta^{n-1}$ is given by the formula (4.1), in which  $a_1$,
$r^0(a_1)$, $r^0(a)$ are replaced by $a_n$, $r^{n-1}(a_1,\dots,a_n)$,
$r^{n-1}(a_1,\dots,a_{n-1},a).$

If we set
$$r^0(a) = R(a) \equiv  \exp(a x + a^{-1} y), $$
we get
$$ \Theta^1(R(a_1),R(a)) =
\frac{2(a a^2_1 - a_1 a^2) + a^3 - a^3_1}{a^3 + a^3_1}R(a_1)R(a) = $$
$$= \sigma(a_1,a)R(a_1)R(a) ,$$
where  the function $\sigma(a_1,a)$  determined by
 $$\sigma(a_1,a) = (a - a_1)/(a_1 + a).$$
Thus, the corresponding function  $r^1(a_1,a)$ is of the form
$\sigma(a_1,a)R(a).$ It is now easy to see that
$$r^n(a_1,\dots,a_n,a) =
R(a)\prod \limits_{i=1}^{n} \sigma(a_i,a). \eqno(4.3)$$

It turns out that $N$-soliton solution of the Tzitz\'{e}ica equation can
be expressed in terms of the function $R$. Obviously, the function
$\tau^0 = 1$ is a solution of (2.2) and the corresponding potential $v^0$
is equal to 1. In this case, the system (1.4) is of the form
$$r_{xy} = r,\quad r_{xx} = a^3 r_y,\quad r_{yy} = a^{-3} r_x .
\eqno(4.4) $$
The general solution of the system (4.4) is
$$ r^0 = c_0 R(a) + c_1 R(a\nu ) + c_2 R(a\nu^2), $$
where $c_i$ are arbitrary constants, $\nu = -(1 + i\sqrt{3})/2.$

Now we will seek expression for $\tau^n(a_1,...,a_n)$. Since
$$\tau_1(a_1) = r^0(a_1)\tau_0 = r^0(a_1),$$
the function $\tau^n(a_1,...,a_n)$ takes the form
$$ r^0(a_1)\wedge\cdots\wedge r^0(a_n) .$$
According to the Theorem 2, the map $r^0(a_1)\wedge\cdots\wedge r^0(a_n)$ is
multilinear. Therefore, it is useful to calculate the expression
$$ R(a_1)\wedge\cdots\wedge R(a_n).$$
From (3.14) and (4.3) we obtain
$$   R(a_1)\wedge\cdots\wedge R(a_n) =
R(a_1)\cdots R(a_n)\prod \limits_{1 \leq i < j \leq n} \sigma(a_i,a_j).
\eqno(4.5)$$
{\bf Lemma 4.} The two-soliton solution $1 - 2(\ln \tau_2)_{xy}$ of the
Tzitz\'{e}ica equation coincides with the potential $1 - 2(\ln \eta_2)_{xy},$
where  $\tau_2$ is given by (2.3) and
$$\eta_2 = (R(a_1) + c_1 R(a_1 \nu))\wedge (R(a_2) + c_2 R(a_2 \nu)).$$
{\it Proof.} Since $\eta_2$ is a multilinear map, the relation (4.5) implies
that
$$R(a_1)\wedge R(a_2) + c_1 R(a_1 \nu)\wedge R(a_2) +
 c_2 R(a_1)\wedge R(a_2 \nu) + c_1 c_2 R(a_1 \nu)\wedge R(a_2 \nu) =  $$
$$= \sigma(a_1,a_2)R(a_1)R(a_2) + c_1 \sigma(a_1 \nu,a_2) R(a_1 \nu) R(a_2) +
$$
$$+c_2\sigma(a_1,a_2 \nu) R(a_1) R(a_2 \nu) +
c_1 c_2\sigma(a_1 \nu,a_2 \nu) R(a_1 \nu)R(a_2 \nu) =$$
$$= \sigma(a_1,a_2)R(a_1)R(a_2) \biggl( 1 +
c_1 \frac{\sigma(a_1 \nu,a_2)}{\sigma(a_1,a_2)}
\frac{R(a_1 \nu)}{R(a_1)} \quad +$$
$$
+\quad c_2 \frac{\sigma(a_1,a_2 \nu)}{\sigma(a_1,a_2)}
 \frac{R(a_2 \nu)}{R(a_2)} +  c_1 c_2 \frac{R(a_1 \nu)R(a_2 \nu)}
{R(a_1)R(a_2)} \biggr).$$

Let
$$b_1 = c_1 \sigma(a_1 \nu,a_2)/\sigma(a_1,a_2),\quad
b_2 = c_2 \sigma(a_1,a_2 \nu)/\sigma(a_1,a_2),$$
$$ A = \sigma(a_1,a_2)R(a_1)R(a_2),$$
then $\eta_2$  is of the form
$$A \biggl( 1 + b_1 \frac {R(a_1 \nu)}{R(a_1)} +
b_2 \frac {R(a_2 \nu)}{R(a_2)} +  b_1 b_2 \frac{\sigma(a_1,a_2)
\sigma(a_1\nu,a_2\nu)
R(a_1 \nu)R(a_2 \nu)}
{\sigma(a_1,\nu a_2)\sigma(\nu a_1,a_2)
R(a_1)R(a_2)} \biggr).$$
Setting
$$a_i = k_i/(\nu - 1), \quad i=1,2,$$
it is easy to check that
$$p_{12} = \frac{\sigma(a_1,a_2)\sigma(a_1\nu,a_2\nu)}
{\sigma(a_1,\nu a_2)\sigma(\nu a_1,a_2)},$$
where $p_{12}$ is given by (2.5). Thus, $\eta_2 = A\tau_2.$
It remains to note that $(\ln A)_{xy} = 0.$

{\bf Remark:} It is easy to see that one-soliton solution
$$v_1 = 1 - 2(\ln (1 + c_1\exp (kx + 3y/k ))_{xy}$$
coincides with the function $1 - 2(\ln \eta_1 )_{xy},$ where
$\eta_1 = R(k/(\nu -1)) + c_1 R(k\nu/(\nu -1)).$

Let us consider the function
$$\eta_n = \biggl(
 R(a_1) + c_1 R(a_1\nu)\biggr) \wedge\cdots\wedge\biggl( R(a_n) +
 c_n R(a_n\nu) \biggr), \eqno(4.6)$$
where $c_i$ are arbitrary constants. It turns out that the $\eta_n$ and
$\tau_n$ ( given by (2.6) ) differ only by a multiplicative exponential
function. The proof of this statement can be found below.

Let $A_0(a_1,\dots,a_n)$ denote the product
$ \prod \limits_{1 \leq i < j \leq n} \sigma(a_i,a_j)$, $A_0(a_I)
= A_{i_1\cdots i_m}$, where
$a_I = (a_1,\cdots,a_{i_1}\nu,a_{i_1+1},\cdots,a_{i_m}\nu,\cdots,a_n).$
Denote by $R^{i_1\cdots i_m}$,  $R_{i_1\cdots i_m}$ the following expressions
$$R(a_1)\wedge\cdots\wedge R(a_{i_1})\wedge R(a_{i_1+1})\wedge \cdots\wedge
R(a_{i_m})\cdots \wedge R(a_n),$$
$$R(a_1)\cdots R(a_{i_1})R(a_{i_1+1})\cdots R(a_{i_m})\cdots R(a_n).$$

\noindent
{\bf Theorem 3.} The function $1 - 2(\ln \tau_n)_{xy}$ is a solution of the
Tzitz\'{e}ica equation.\\
{\it Proof.} Since $\eta_n$ is a multilinear map, the relations (4.5) and
(4.6) imply
$$R(a_1)\wedge\cdots\wedge R(a_n) +
\sum\limits _{i=1}^n c_i R(a_1)\wedge\cdots R(a_i\nu)\cdots\wedge R(a_n) + $$
$$+ \sum\limits _{m=2}^n \sum\limits _{1\leq i_1<\cdots <i_m\leq n}
c_{i_1}\cdots c_{i_m} R^{i_1\cdots i_m} = A_0 R(a_1)\cdots R(a_n) +
 \sum\limits _{i=1}^n c_i A_i R_i + $$
$$+ \sum\limits _{m=2}^n \sum\limits _{1\leq i_1<\cdots <i_m\leq n}
c_{i_1}\cdots c_{i_m} A_{i_1\cdots i_m} R_{i_1\cdots i_m} =
A_0 R(a_1)\cdots R(a_n) \omega_n,$$
where  $\omega_n$ means
$$ 1 + \sum\limits_{i=1}^n c_i \frac{A_i R(a_i\nu)}{A_0 R(a_i)} +
\sum\limits _{m=2}^n \sum\limits_{1\leq i_1<\cdots <i_m\leq n}
c_{i_1}\cdots c_{i_m}
\frac{ A_{i_1\cdots i_m}R(a_{i_1}\nu)\cdots R(a_{i_m}\nu)}
{ A_0 R(a_{i_1})\cdots R(a_{i_m}) }. $$
It is easy to see that $(\ln A_0 R(a_1)\cdots R(a_n) )_{xy} = 0$. Thus,
it suffices to prove that $\omega_n$ is equal to $\tau_n$.

Setting
$$b_i = \frac {c_i A_i}{A_0},\quad  p_{i_1\cdots i_m} =
\frac{A_0^{m-1} A_{i_1\cdots i_m} }{A_{i_1}\cdots A_{i_m} },$$
we can rewrite $\omega_n$ in the form
$$1 + \sum\limits_{i=1}^n b_i \frac{R(a_i\nu)}{R(a_i)} +
\sum\limits _{m=2}^n \sum\limits_{1\leq i_1<\cdots <i_m\leq n}
b_{i_1}\cdots b_{i_m} p_{i_1\cdots i_m}
\frac { R(a_{i_1}\nu)\cdots R(a_{i_m}\nu)}{R(a_{i_1})\cdots R(a_{i_m})}.$$
Now it is convenient to express $A_i, A_{i_1\cdots i_m}, p_{i_1\cdots i_m}$
in terms of the function $\sigma.$ According to the definition, $A_{i_1}$
can be written as
$$ \frac{\sigma(a_1,a_{i_1}\nu)\cdots \sigma(a_{i_1-1},a_{i_1}\nu)
\sigma(a_{i_1}\nu,a_{i_1+1})\cdots \sigma(a_{i_1}\nu,a_n)
 \prod \limits_{1 \leq i < j \leq n} \sigma(a_i,a_j)}
{ \sigma(a_1,a_{i_1})\cdots  \sigma(a_{i_1},a_n)} =$$
$$ = \biggl( \prod \limits_{l=1}^{i_1-1} \frac{\sigma(a_l,a_{i_1}\nu)}
{\sigma(a_l,a_{i_1})}
    \prod \limits_{l=i_1+1}^{n} \frac{\sigma(a_{i_1}\nu,a_l)}
{\sigma(a_{i_1},a_l)}\biggr) A_0. $$
Similarly, $A_{i_1 i_2}$ and $A_{i_1\cdots i_m}$ are given by
$$A_{i_1 i_2} =
 \frac{\sigma(a_{i_1},a_{i_2})\sigma(a_{i_1}\nu,a_{i_2}\nu)}
{\sigma(a_{i_1}\nu,a_{i_2})\sigma(a_{i_1},a_{i_2}\nu)}
 \biggl(
 \prod \limits_{l=1}^{i_1-1}
 \frac{\sigma(a_l,a_{i_1}\nu)}{\sigma(a_l,a_{i_1})}
 \prod \limits_{l=i_1+1}^{n} \frac{\sigma(a_{i_1}\nu,a_l)}
{\sigma(a_{i_1},a_l)}\biggr) \times $$
$$ \times
 \biggl(
 \prod \limits_{l=1}^{i_2-1}
 \frac{\sigma(a_l,a_{i_2}\nu)}{\sigma(a_l,a_{i_2})}
 \prod \limits_{l=i_2+1}^{n} \frac{\sigma(a_{i_2}\nu,a_l)}
{\sigma(a_{i_2},a_l)} \biggr)A_0,$$
$$A_{i_1\cdots i_m} =
 \prod \limits_{1\leq j<k\leq m}  \biggl[
 \frac{\sigma(a_{i_j},a_{i_k})\sigma(a_{i_j}\nu,a_{i_k}\nu)}
{\sigma(a_{i_j}\nu,a_{i_k})\sigma(a_{i_j},a_{i_k}\nu)}
 \biggl(
 \prod \limits_{l=1}^{i_j-1}
 \frac{\sigma(a_l,a_{i_j}\nu)}{\sigma(a_l,a_{i_j})}
 \prod \limits_{l=i_j+1}^{n} \frac{\sigma(a_{i_j}\nu,a_l)}
{\sigma(a_{i_j},a_l)}\biggr) \times $$
$$ \times
 \biggl(
 \prod \limits_{l=1}^{i_k-1}
 \frac{\sigma(a_l,a_{i_k}\nu)}{\sigma(a_l,a_{i_k})}
 \prod \limits_{l=i_k+1}^{n} \frac{\sigma(a_{i_k}\nu,a_l)}
{\sigma(a_{i_k},a_l)} \biggr) \biggr]A_0.$$
This yields
$$p_{i_1 i_2} \equiv \frac{A_0 A_{i_1 i_2}}{A_{i_1}A_{i_2}} =
\frac{\sigma(a_{i_1},a_{i_2})\sigma(a_{i_1}\nu,a_{i_2}\nu)}
{\sigma(a_{i_1}\nu,a_{i_2})
\sigma(a_{i_1},a_{i_2}\nu)}, $$
$$p_{i_1\dots i_m} = \prod \limits_{1\leq j<k\leq m}
 \frac{\sigma(a_{i_j},a_{i_k})\sigma(a_{i_j}\nu,a_{i_k}\nu)}
{\sigma(a_{i_j}\nu,a_{i_k})\sigma(a_{i_j},a_{i_k}\nu)} =
\prod \limits_{1\leq j<k\leq m} p_{i_j i_k}.$$
Setting  $a_i = k_i/(\nu -1)$ and using the last representation of
$p_{i_1\dots i_m}$, we can conclude that $\omega_n = \tau_n$.

\vspace{10mm}
{\bf  ACKNOWLEDGEMENTS}
\nopagebreak[4]

We wish to thank Sergey P.Tsarev for useful discussions relating to this
work. Besides, he acquainted us with Tzitz\'{e}ica's articles \cite{tz1,tz2}.
This work was supported in part by Russian Foundation for Basic Research
through grant 96-01-00047.

\end{document}